\documentclass[aps,twocolumn,amsmath,amssymb,showpacs]{revtex4-1}
\usepackage{graphicx,color,hyperref}
\begin{document}
\def\be{\begin{equation}}
\def\ee{\end{equation}}
\def\bea{\begin{eqnarray}}
\def\eea{\end{eqnarray}}
\def\bef{\begin{figure}[h!]}
\def\eef{\end{figure}}

\def\a{\alpha}
\def\th{\theta}
\def\eps{\epsilon}
\def\fr{\frac}
\def\l{\label}

\title{Overdamped dynamics of long-range systems on a one-dimensional
lattice: Dominance of the mean-field mode and phase transition}
\author{Shamik Gupta$^1$}
\email{shamikg1@gmail.com}
\author{Alessandro Campa$^2$}
\email{alessandro.campa@iss.infn.it}
\author{Stefano Ruffo$^{1,3}$}
\email{stefano.ruffo@gmail.com}
\affiliation{$^1$Laboratoire de Physique de l'\'{E}cole Normale Sup\'{e}rieure de Lyon, Universit\'{e}
de Lyon, CNRS, 46 All\'{e}e d'Italie, 69364 Lyon c\'{e}dex 07, France \\
$^2$Complex Systems and Theoretical Physics Unit, Health and Technology
Department, Istituto Superiore di Sanit\`{a}, and INFN Roma1, Gruppo
Collegato Sanit\`{a}, Viale Regina Elena 299,
00161 Roma, Italy \\
$^3$Dipartimento di Energetica ``Sergio Stecco'' and CSDC,
Universit\`{a} di Firenze, CNISM and INFN, via S. Marta 3, 50139 Firenze, Italy}
\begin{abstract}
We consider the overdamped dynamics of a paradigmatic long-range
system of particles residing on the sites of a one-dimensional lattice,
in the presence of thermal noise.
The internal degree of freedom of each particle is a periodic variable which is coupled to
those of other particles with an attractive $XY$-like interaction. The
coupling strength decays with the interparticle separation $r$ in space as
$1/r^\alpha$; ~$0 < \alpha < 1$. We study the dynamics of the model in the continuum limit by
considering the Fokker-Planck equation for the evolution of the spatial density
of particles. We show that the equation allows a linearly
stable stationary state which is always uniform in space, being non-uniform in
the internal degrees below a critical 
temperature $T=1/2$ and uniform above, with a phase transition
between the two at $T=1/2$. The state is the same as the
equilibrium state of the mean-field version of the model, obtained by
considering $\alpha=0$. Our analysis also lets us to compute the
growth and decay rates of spatial Fourier modes of density fluctuations.
The growth rates compare very well with numerical simulations.

\end{abstract}
\date{\today}
\pacs{05.20.-y, 05.40.-a, 05.70.Fh}
\maketitle
\section{Introduction}
In long-range interacting systems, the interparticle potential in $d$
dimensions decays at large separation, $r$, as
$1/r^{\alpha}$, with $0 \le \alpha \le d$
\cite{review1,review2,review3}. Examples are gravitational systems
\cite{Chavanis:2006}, plasmas \cite{Escande:2010}, two-dimensional
hydrodynamics \cite{Bouchet:2012}, charged and dipolar systems
\cite{Bramwell:2010}, etc. Long-range systems are non-additive, and have equilibrium properties unusual for short-range systems, e.g., inequivalence of statistical
ensembles \cite{Barre:2001}.
Besides, they show intriguing dynamical features, e.g., broken
ergodicity \cite{Mukamel:2005} and slow relaxation towards equilibrium
\cite{Mukamel:2005,Yamaguchi:2004}.

A paradigmatic example of long-range systems is the so-called
Hamiltonian mean-field (HMF) model, comprising a system of $N$ interacting particles
evolving under Hamilton dynamics \cite{Ruffo:1995}. The $i$th particle, $i=1,2,\ldots,N$, has an internal degree of freedom $\th_i$, a
periodic variable of period $2\pi$. The $\th_i$'s are coupled to one
another via an attractive $XY$-like interaction $\sim -\cos(\th_i-\th_j)$ between the $i$th and $j$th particles, with the
coupling strength being of the ``mean-field" type: It is equal for every
pair of particles. As a result, the model is defined without requiring
any lattice structure for the particles to reside on. In this
model, a wide class of initial distributions relaxes to the
Boltzmann-Gibbs equilibrium state over times that diverge with
the system size \cite{Yamaguchi:2004}. As a consequence, the system
in the thermodynamic limit does not ever attain equilibrium, instead it remains trapped in intermediate quasistationary
states. Besides the HMF model, these states have been
observed in a wide variety of long-range systems ranging from models with
spin dynamics \cite{Gupta:2011} and plasma physics \cite{Levin:2008} to one-dimensional and two-dimensional gravity
\cite{Joyce:2011,Levin:2010}. The HMF model is a
prototype for systems with unscreened long-range interactions
leading to long-lived spatially inhomogeneous states. Moreover, the model does
not have a singularity at short length scales so that one can
exclusively focus on effects of long-range interactions, in a way
similar to studies of gravitational systems where a short-range
softening is introduced \cite{Binney}. 

To explore how far the results obtained within the
HMF model extend beyond its setting of mean-field coupling, the model
has been considered on a one-dimensional periodic lattice on the sites of
which the particles reside. Particles on different sites are coupled as in the HMF model,
with the difference that the coupling strength decays with their
separation $r$ along the lattice
length as $1/r^\alpha$, with $0 < \alpha < 1$. The resulting model is
known as the $\alpha$-HMF model \cite{Anteneodo:1998}. {\bf The HMF model is
recovered on considering $\alpha=0$.}
Within a canonical ensemble, the $\alpha$-HMF model has the same equilibrium state as those of the HMF model \cite{Campa:2000-2003}.
There have been many studies dealing with various dynamical aspects of the $\alpha$-HMF
model that have illustrated their difference from those of the HMF
model, e.g., Lyapunov exponents \cite{ineq1,ineq2}, and others \cite{ineq3,ineq4}. Hence, the equivalence of
the equilibrium state of the two models raises an immediate
question: How does the dynamics of the $\alpha$-HMF
model lead to the same equilibrium state as that of its mean-field
counterpart, the HMF model? This work is a step towards
answering this question regarding the mean-field dominance in dictating the
equilibrium state.

To address the issue, let us consider the simple setting of the
$\alpha$-HMF model evolving in
the presence of thermal noise and dissipation. The $i$th particle has
the equations of motion
\bea
&&\fr{d\th_i}{dt}=p_i, \label{alphaHMF-EOM1} \\ 
&&\fr{dp_i}{dt}=-\gamma p_i+\frac{1}{\widetilde{N}}\sum_{j=1,j \ne
i}^{N}\fr{\sin(\th_j-\th_i)}{(|j-i|_{c})^{\a}}+\sqrt{\gamma} \eta_i(t),
\label{alphaHMF-EOM2}
\eea
where $p_i$ denotes the momentum of the $i$th particle, while $\gamma$ is the
damping constant. The second term on the right hand side
of Eq. (\ref{alphaHMF-EOM2}) is
the force on the $i$th particle arising from the interaction potential: 
The quantity $|j-i|_{c}$ is the closest distance on the periodic lattice between the $i$th and $j$th sites:
\be
|j-i|_c={\rm min}(|j-i|,N-|j-i|), 
\ee
while
\be
\widetilde{N}=\sum_{j=1}^N(|j-i|_c)^{-\alpha} ~\forall ~i.
\ee
From Eq. (\ref{alphaHMF-EOM2}), since $\a < 1$, the cumulative interaction of one particle with all the
remaining particles with aligned $\theta$'s would diverge in the limit $N
\to \infty$ in the absence of the normalization $\widetilde{N}$, which
explains its inclusion \cite{Campa:2000-2003}.
In Eq.  (\ref{alphaHMF-EOM2}), $\eta_i(t)$ is a Gaussian white noise:
\bea
&&\langle \eta_i(t) \rangle=0, \\
&&\langle \eta_i(t) \eta_j(t') \rangle=2T\delta_{ij}\delta(t-t'),
\eea
where $T$ denotes the temperature, while the angular brackets denote averaging over noise realizations.
In this work, we take the Boltzmann constant to be unity. The
equations of motion (\ref{alphaHMF-EOM1}) and
(\ref{alphaHMF-EOM2}) describe the evolution of the $\alpha$-HMF model
within a canonical ensemble. We note that for $\alpha=0$, these equations reduce to
those of the Brownian mean-field model considered in Refs. \cite{Acebron:1998-2000} and \cite{Chavanis:2005}.

Here, we will study the overdamped limit of the
equations of motion, (\ref{alphaHMF-EOM1}) and (\ref{alphaHMF-EOM2});
after a time rescaling, these equations then reduce to the following
Langevin equation for the $i$th particle:  
\be
\fr{d\th_{i}}{dt}=\frac{1}{\widetilde{N}}\sum_{j=1,j \ne
i}^{N}\fr{\sin(\th_j-\th_i)}{(|j-i|_{c})^{\a}}+\eta_i(t).
\l{EOM}
\ee
We analyze the dynamics (\ref{EOM}) in the continuum limit $N \to
\infty$, when the lattice becomes a continuous segment characterized by
the spatial coordinate $s \in [0,1]$. In this limit, the local density
of oscillators $\rho(\th; s,t)$  evolves in time following a Fokker-Planck equation. 

We show that the Fokker-Planck equation allows a
stationary state which is uniform in both $\th$ and $s$. By performing a linear stability analysis
of such a state, we find that when it is unstable, different
spatial Fourier modes of fluctuations have different critical
temperatures below which the modes grow exponentially in time with
different rates. The largest critical temperature, $T_{c,0}=1/2$,
corresponds to the spatially independent zero Fourier mode, i.e. the
mean-field mode. Above this temperature, all the Fourier modes decay in
time, thereby stabilizing the uniform state. Below $T_{c,0}$, our
numerical simulations starting from the uniform state show that the
unstable non-zero Fourier modes, growing exponentially in time at short
times, nevertheless decay at
long times to zero. By contrast, the mean-field mode grows and
reaches a non-zero value, corresponding to a non-uniform stationary
state, i.e. a state which is uniform in $s$, but non-uniform in $\theta$. We perform a linear stability analysis of the non-uniform state
to demonstrate that indeed at temperatures below $T_{c,0}$, all the
Fourier modes of fluctuations decay in time, thereby stabilizing the
non-uniform state.

The long time dominance of the mean-field mode leading to a
stationary state always uniform in space, here observed
in the alpha-HMF model evolving under the canonical and overdamped
dynamics, has also been observed when the model evolves within a
microcanonical ensemble. In the latter case, the dynamical equations are
obtained from Eqs. (\ref{alphaHMF-EOM1}) and
(\ref{alphaHMF-EOM2}) by substituting $\gamma=0$ \cite{Bachelard:2011}.
In the case of microcanonical dynamics, this dominance was just a numerical observation with no theoretical
justification, while within the canonical and overdamped dynamics, we
show how the mean-field mode arises as a linearly stable stationary solution of
the Fokker-Planck equation.

The paper is structured as follows. In section \ref{uniform}, we introduce
the continuum limit of the dynamics (\ref{EOM}), and perform the linear
stability analysis of the uniform stationary state. In section
\ref{numerics}, we present numerical simulations in support of the
analysis, in particular, to show the agreement of the growth rates of
the unstable modes, the decay at long times of the unstable non-zero modes at all
temperatures $T <1/2$, and the associated dominance of the mean-field mode. In section \ref{nonuniform}, we perform the linear
stability analysis of the non-uniform stationary state. We end the paper
with conclusions and perspectives.
\section{Continuum limit: Uniform stationary state and dynamics of Fourier modes}
\l{uniform}
In the continuum limit $N \to \infty$, the lattice of the system (\ref{EOM}) is densely filled with sites.
Let us introduce the variable $s=i/N$ to denote the spatial
coordinate along the lattice length, such that in the continuum limit
it becomes the continuous variable $s \in [0,1]$. In this limit, we define a local density of
particles $\rho(\th;s,t)$ such that of all particles located between $s$ and $s+ds$ at time
$t$, the fraction $\rho(\th;s,t)d\th$ have their degrees of freedom between $\th$ and $\th+d\th$. This
density is non-negative, $2\pi$-periodic in $\th$, and obeys the
normalization
\be
\int_0^{2\pi}d\th ~\rho(\th;s,t)=1 ~~\forall ~s.
\l{rho-normalization}
\ee
In the continuum limit, the equation of motion is given by
\bea
\fr{\partial \theta(s,t)}{\partial t}=\kappa(\a)\int
d\th'ds'\fr{\sin(\th'-\th)}{(|s'-s|_{c})^{\alpha}}\rho(\th';s',t)+\eta(s,t),
\nonumber \\
\l{EOMcontinuum}
\eea
where 
\be
\kappa(\a)^{-1}=\int_{s-1/2}^{s+1/2}\fr{ds'}{(|s'-s|_c)^\a},
\l{kappa}
\ee
and
\be
|s'-s|_c={\rm min}(|s'-s|,1-|s'-s|). 
\ee
Also, we have
\bea
&&\langle \eta(s,t)\rangle=0, \\
&&\langle \eta(s,t)\eta(s',t') \rangle=2T\delta(s-s')\delta(t-t').
\eea

The time evolution of $\rho(\th; s,t)$ follows the Fokker-Planck equation
\bea
\fr{\partial\rho}{\partial
t}&=&-\kappa(\a)\fr{\partial}{\partial \th}\Big[\Big(\int
d\th'ds'\fr{\sin(\th'-\th)}{(|s'-s|_c)^{\alpha}}\rho(\th';s',t)\Big)\rho\Big]\nonumber \\
&&+T\fr{\partial^2\rho}{\partial \th^2}.
\l{FP-equation}
\eea
Its stationary solution, obtained by setting
the left hand side to zero, is
\be
\rho_0(\th;s) \propto \exp\Big[\fr{\kappa(\a)}{T}\int
d\th'ds'\fr{\cos(\th'-\th)}{(|s'-s|_c)^{\alpha}}\rho_0(\th';s')\Big].
\l{nonuniform-rho}
\ee

Consider the particular stationary state which is uniform both in $\th$ and in the
spatial coordinate $s$,
\be
\rho_{0}=\fr{1}{2\pi}.
\l{uniform-state}
\ee
This state, although stationary, might be destabilized by the thermal
fluctuations inherent in the dynamics. Let us then analyze the
stability, in particular, linear stability of the state with respect to small
thermal fluctuations $\delta\rho(\th;s,t)$. To this end, we write
\be
\rho(\th;s,t)=\fr{1}{2\pi}+\delta\rho(\th;s,t);
~~\delta\rho(\th;s,t) \ll 1.
\l{delta-rho-definition}
\ee
Substituting Eq. (\ref{delta-rho-definition}) into Eq. (\ref{FP-equation}) and keeping terms to
linear order in $\delta \rho$, we find that $\delta \rho$ satisfies
\bea
\fr{\partial\delta\rho}{\partial
t}&=&\fr{\kappa(\alpha)}{2\pi}\int d\th'ds'
~\fr{\cos(\th'-\th)}{(|s'-s|_c)^{\alpha}}\delta\rho(\th';s',t)+T\fr{\partial^2
\delta\rho}{\partial \th^2}. \nonumber \\
\l{linearized-equation}
\eea
Expressing $\delta \rho$ in terms of its Fourier series with respect
to the periodic variable $\th$ as
\be
\delta\rho(\th;s,t)=
\sum_{k=-\infty}^\infty\widehat{\delta\rho}_k(s,t) e^{ik\th},
\l{delta-rho-definition1}
\ee
we find from Eq. (\ref{linearized-equation}) that only the modes $k=\pm
1$ are affected by the coupling between the particles, and that $\widehat{\delta \rho}_{\pm
1}$ satisfies
\bea
&&\fr{\partial\widehat{\delta\rho}_{\pm1}}{\partial
t}=\fr{\kappa(\a)}{2}\int ds'
\fr{\widehat{\delta\rho}_{\pm1}(s',t)}{(|s'-s|_c)^{\a}}-T\widehat{\delta\rho}_{\pm1}. 
\l{linearized-equation-1}
\eea
One thus gets a set of equations for each position $s$, all coupled
together by the second term on the right hand side of Eq.
(\ref{linearized-equation-1}). For $k \ne \pm 1$, one has
\bea
&&\fr{\partial\widehat{\delta\rho}_{\pm k}}{\partial
t}=-T\widehat{\delta\rho}_{\pm k}; ~~k \ne 1. 
\eea
Thus, $\widehat{\delta\rho}_{\pm k}(s,t)$ for $k \ne 1$
decays exponentially in time as
$\exp(-Tt)$, so that these modes cannot destabilize the uniform state
(\ref{uniform-state}).

Since we have a periodic lattice, to solve Eq.
(\ref{linearized-equation-1}), we next consider the spatial
Fourier series of $\widehat{\delta \rho}_{\pm 1}(s,t)$:
\be
\widehat{\delta \rho}_{\pm 1}(s,t)=\sum_{m=-\infty}^\infty \overline{\delta \rho}_{\pm
1,m}(t)e^{i2\pi ms}.
\l{spatial-FT}
\ee
On substituting in Eq. (\ref{linearized-equation-1}), we get 
\bea
&&\fr{\partial\overline{\delta\rho}_{\pm1,m}}{\partial
t}=\fr{\kappa(\a)\Lambda_m(\a)}{2}\overline{\delta\rho}_{\pm1,m}-T\overline{\delta\rho}_{\pm1,m},
\l{linearized-equation-2}
\eea
where
\bea
\Lambda_m(\a)&=&\int_{s-1/2}^{s+1/2} ds' ~\fr{e^{i2\pi
m(s'-s)}}{(|s'-s|_c)^\a}.
\l{lambda}
\eea

Note that $\Lambda_m(\a)=\Lambda_{-m}(\a)$. It is known that
$\Lambda_m(\a) \ge 0$, and that it is a monotonically decreasing
function of $|m|$ when $m$ is an integer; moreover, $\Lambda_m(\a) \to 0$
as $m \to \pm \infty$ \cite{Gupta:2012}. 

Equation (\ref{linearized-equation-2}) has the solution
\bea
\overline{\delta\rho}_{\pm1,m}(t)=
\exp\Big[\Big(\fr{\kappa(\a)\Lambda_m(\a)}{2}-T\Big)t\Big]\overline{\delta\rho}_{\pm1,m}(0).
\eea
It then follows from our linear stability analysis that the $m$th
spatial mode of fluctuations $\overline{\delta\rho}_{\pm1,m}(t)$ either
(i) decays in time, which happens at temperatures such that $T >
\fr{\kappa(\a)\Lambda_m(\a)}{2}$, or (ii) grows in time at temperatures $T < \fr{\kappa(\a)\Lambda_m(\a)}{2}$.
The borderline between these two behaviors is achieved
at the critical temperature $T_{c,m}$ for the $m$th mode, given by 
\be
T_{c,m}=\fr{\kappa(\a) \Lambda_m(\a)}{2}.
\l{Tcm}
\ee
The $m$th mode at $T < T_{c,m}$ grows in time as $\exp[(T_{c,m}-T)t]$. 
With $\Lambda_m(\a)=\Lambda_{-m}(\a)$, the Fourier modes
$m$ and $-m$ have the same behavior in time and the same critical temperature. 
Since $\Lambda_{m}(\a)$ decreases on increasing $|m|$, we conclude that
\be
T_{c,0} > T_{c,1} > T_{c,2} > \ldots.
\l{Tc-hierarchy}
\ee
Note that $T_{c,0}$ is the critical temperature for the $m=0$ mode, i.e., the
``mean-field" mode; we have
\be
T_{c,0}=\fr{\kappa(\a) \Lambda_0(\a)}{2}=\fr{1}{2}.
\ee
Thus, above $T=1/2$, when all the spatial modes decay in time, the uniform
state (\ref{uniform-state}) is stable. On the other hand, the state is
unstable when $T < 1/2$. At temperatures
between $1/2$ and $T_{c,1}$, the $m=0$ mode grows in time, while all
the other modes decay in time. At temperatures
between $T_{c,2}$ and $T_{c,1}$, the $m=0,\pm 1$ modes grow in time, while all
the other modes decay in time, and so on. In Table \ref{table-Tc}, we
show representative numbers for these critical temperatures for
$\alpha=0.25,0.50,0.75$.
\begin{table}[h!]
\begin{tabular}{|c|c|c|c|c|c|}
\hline
 $\alpha$ & $T_{c,1}$ & $T_{c,2}$ & $T_{c,3}$  & $T_{c,4}$ & $T_{c,5}$   \\ \hline \hline 
0.25 & 0.082555 & 0.042070 & 0.033719 & 0.025764 & 0.022664 \\ \hline 
0.50 & 0.186991 & 0.122063 & 0.103418 & 0.087614 & 0.079556 \\ \hline 
0.75 & 0.321783 & 0.262301 & 0.239974 & 0.221807 & 0.210691 \\ \hline 
\end{tabular}
\caption{Critical temperatures $T_{c,m}; m=1,\ldots,5$ for
$\alpha=0.25,0.50,0.75$.
Note that $T_{c,0}=1/2$, independent of $\alpha$. When $\alpha=0$,
one has $T_{c,m}=0$ for $m \ne 0$. When $\alpha$ approaches unity from
below, all the $T_{c,m}$'s for $m \ne 0$ approach $T_{c,0}$.} 
\l{table-Tc}
\end{table}

\section{Comparison with numerical simulations}
\l{numerics}
\begin{figure}[h!]
\includegraphics[width=75mm]{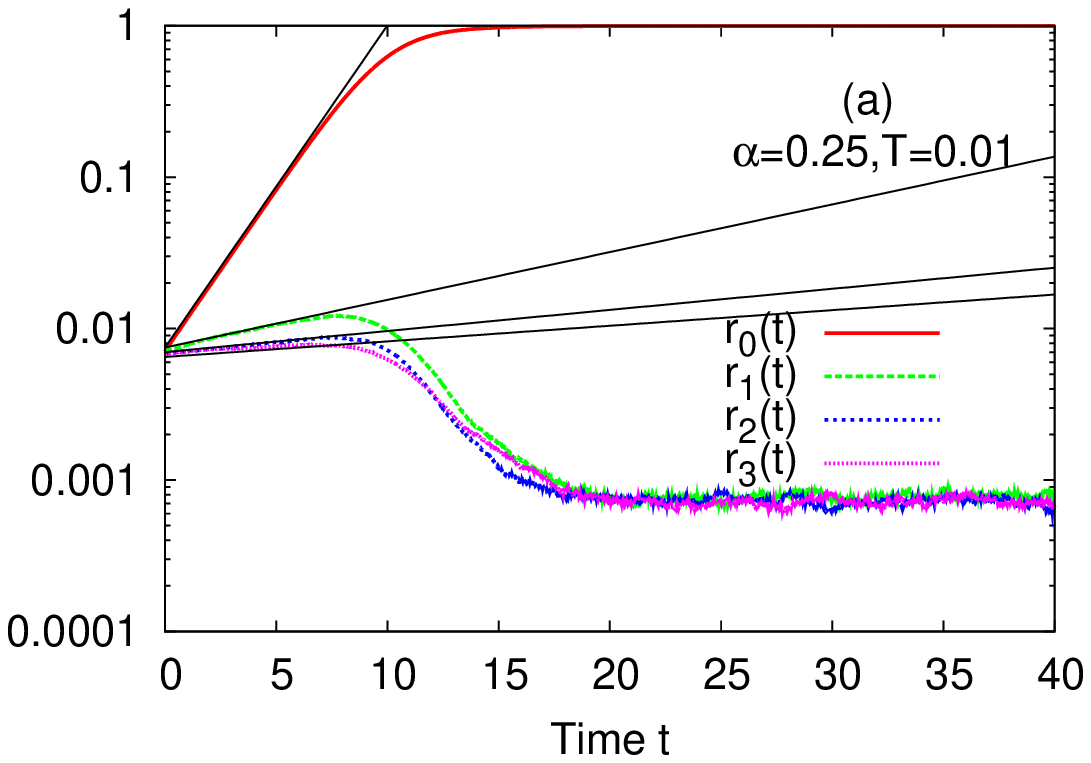}\\ 
\includegraphics[width=75mm]{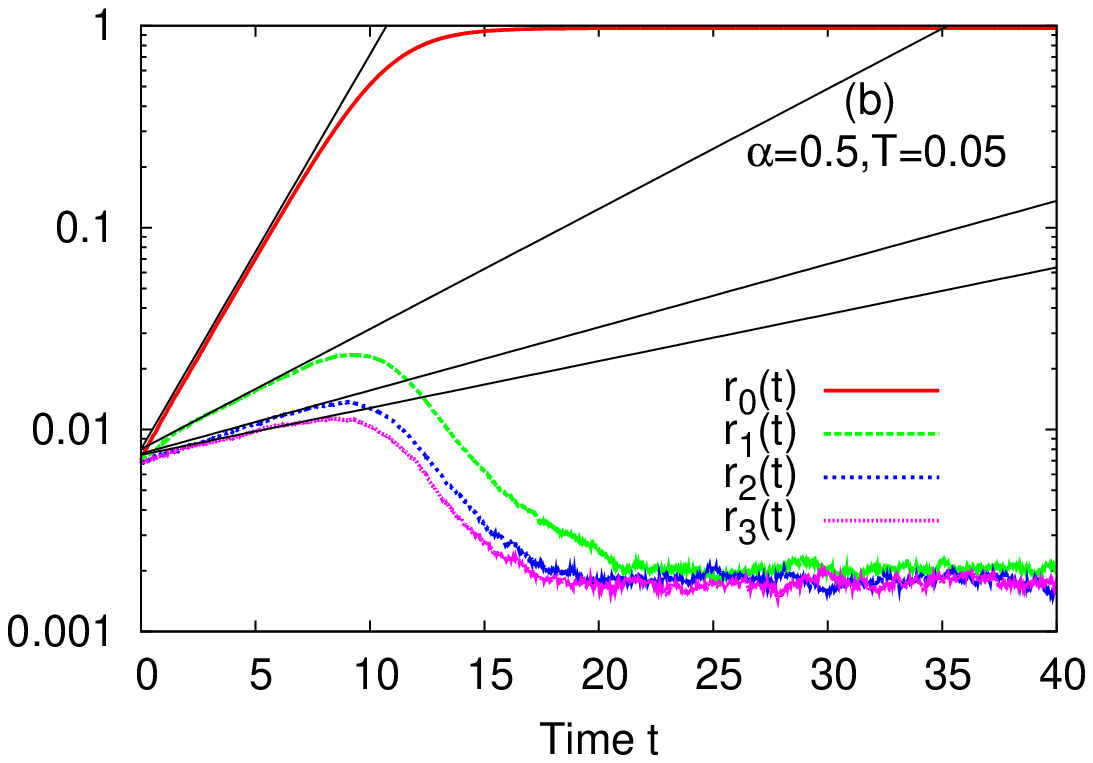}\\
\includegraphics[width=75mm]{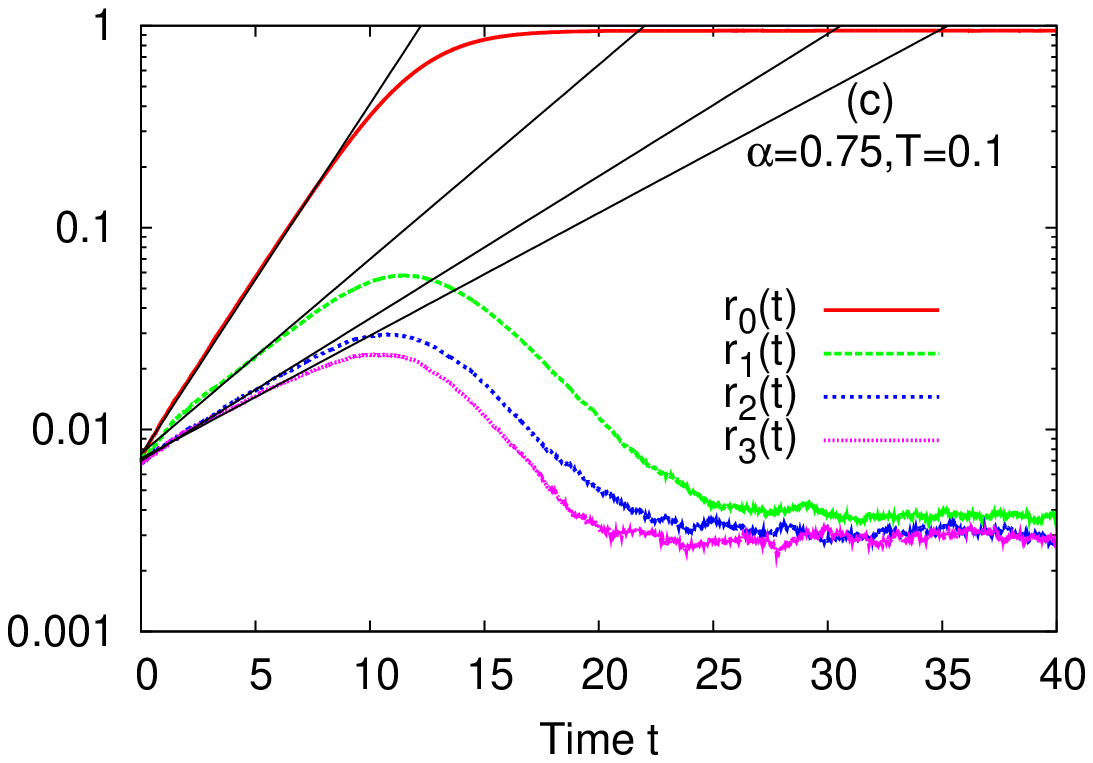}
\caption{(Color online) Time evolution of the observables
$r_0(t),r_1(t),r_2(t)$, and $r_3(t)$ starting from an initial state
$\{\th_i(0); i = 1,2, . . . ,N\}$, where the $\th_i$'s are chosen uniformly in
$[-\pi,\pi]$. Thus, initially, the system is in the uniform state (\ref{uniform-state}). Here, the values of $\alpha$ and
temperature $T$ are (a) $\alpha=0.25, T=0.01$, (b)  $\alpha=0.5,
T=0.05$, and (c) $\alpha=0.75, T=0.1$. From Table (\ref{table-Tc}), it
follows then that the Fourier modes $0, 1, 2, 3$ in particular are linearly
unstable. Consequently, $r_0(t),r_1(t),r_2(t)$, and $r_3(t)$ all grow in
time from their initial values. However, the plots show that in the
long-time limit, $r_0(t)$ attains a value very close to unity, while
$r_1(t),r_2(t),r_3(t)$ all decay to a value very close to zero. The data
in the plots are obtained from
numerical simulations with $N = 2^{14}$, and involve averaging over $100$
independent initial conditions and dynamical realizations. The straight
lines show the initial exponential growth with rates given by $(T_{c,m}-T)$,
where the values of $T_{c,m}$'s can be read off from Table
\ref{table-Tc}. The agreement of the growth rates between theory and simulations is very
good.}
\label{alpha-T-observables}
\end{figure}

\begin{figure}[h!]
\includegraphics[width=75mm]{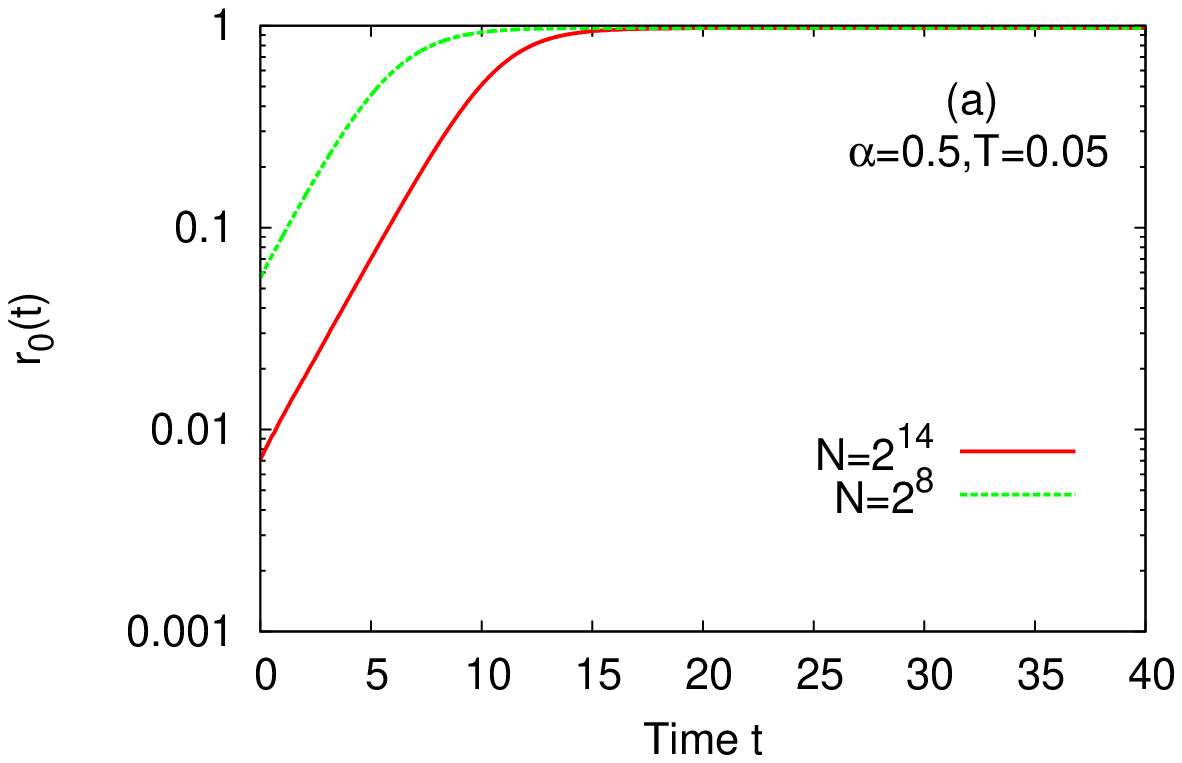}\\ 
\includegraphics[width=75mm]{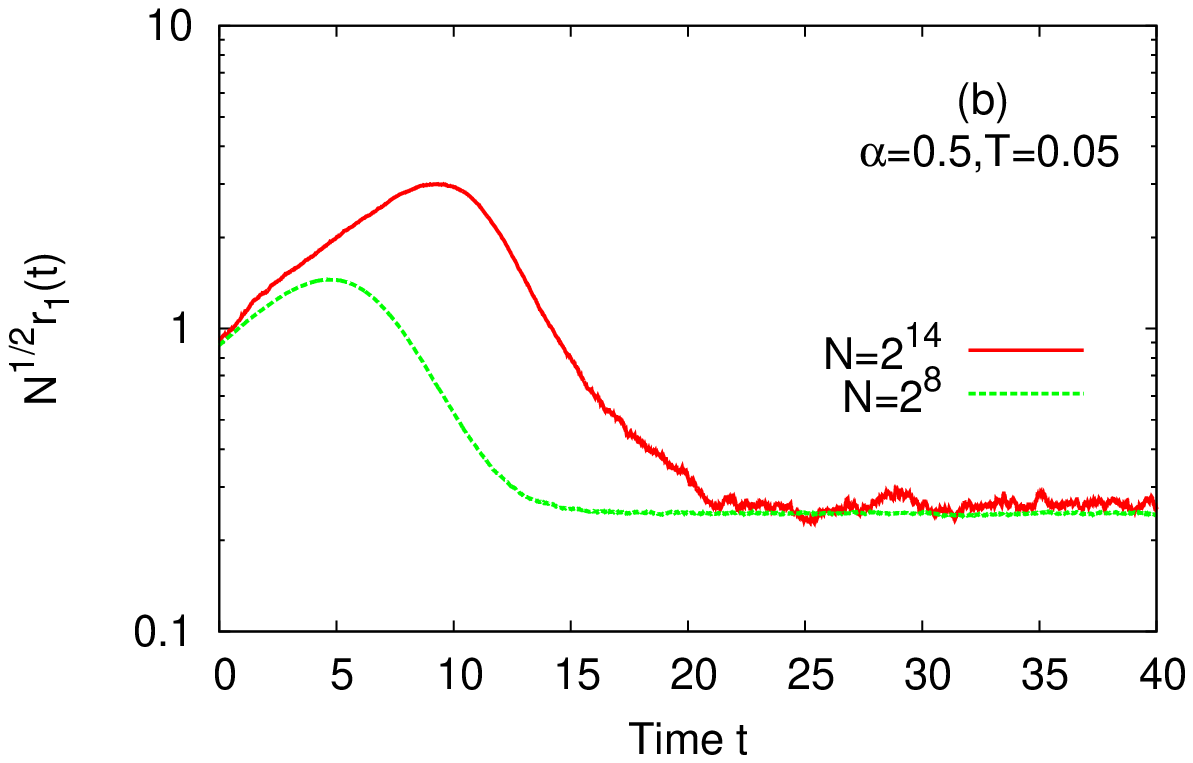}\\
\includegraphics[width=75mm]{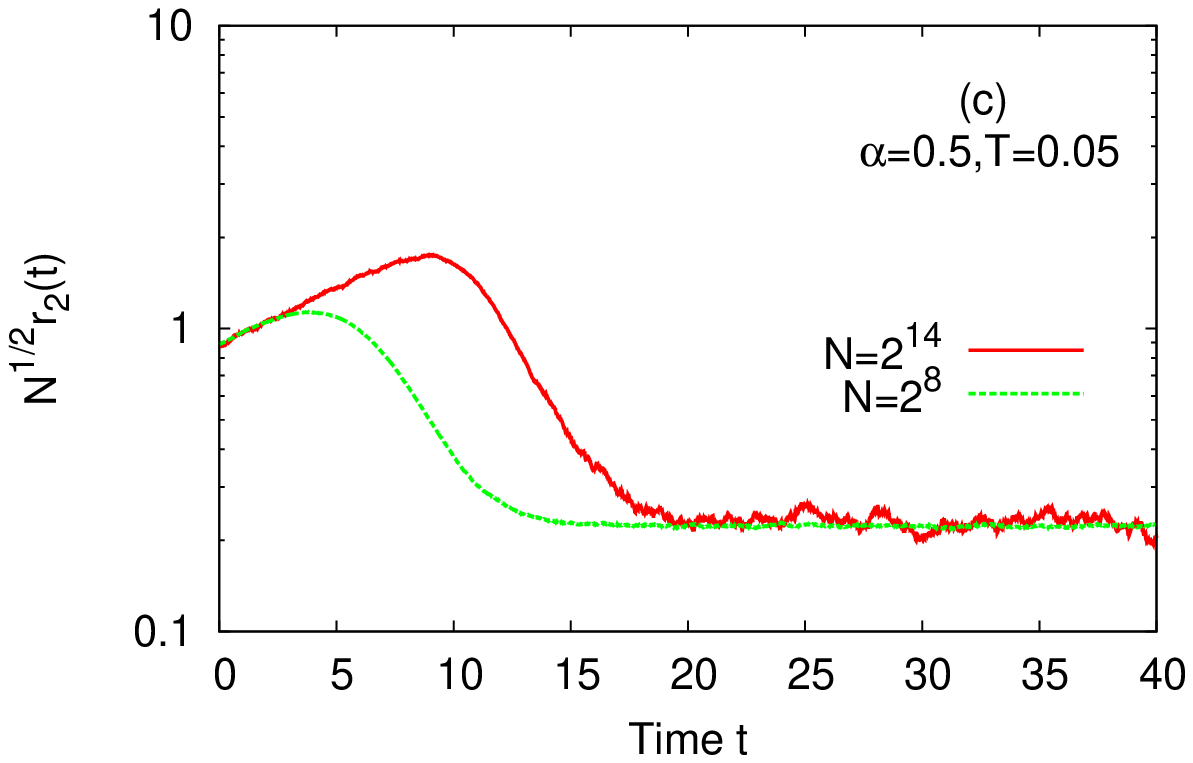}\\
\includegraphics[width=75mm]{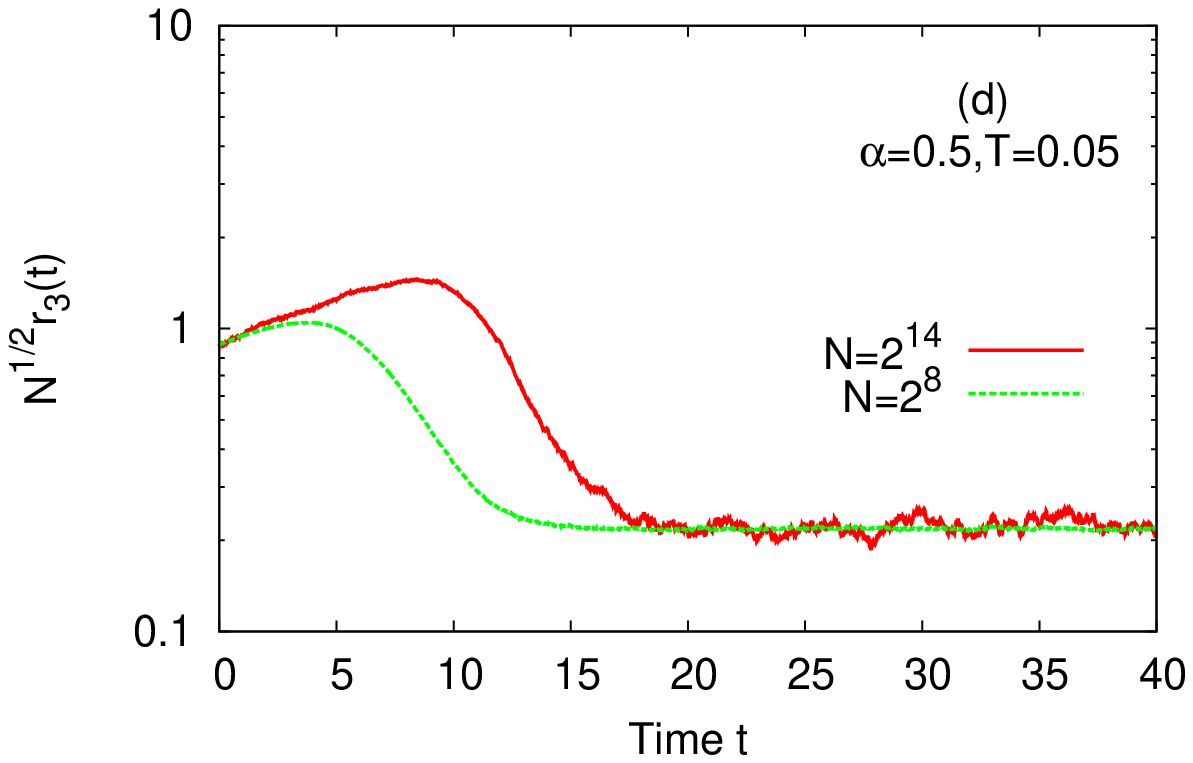}
\caption{(Color online) Time evolution of the observables
$r_0(t),r_1(t),r_2(t)$, and $r_3(t)$ starting from an initial state
$\{\th_i(0); i = 1,2, . . . ,N\}$, where the $\th_i$'s are chosen uniformly in
$[-\pi,\pi]$. Here, $\alpha=0.5$ and $T=0.05$. The plots show that in the
long-time limit, $r_0(t)$ attains a value very close to unity and does
not scale with the system size $N$, while 
$r_1(t),r_2(t),r_3(t)$ all decay to values that scale with $N$ as $1/\sqrt{N}$. The data involve averaging over $100$ 
independent initial conditions and dynamical realizations for
$N=2^{14}$, and over $5000$ realizations for $N=2^8$.}
\label{scaling-fig}
\end{figure}

In order to test the theoretical predictions in the limit $N \to \infty$ obtained in the preceding
section regarding values of
critical temperatures, and growth and decay of Fourier modes depending on the
temperature, we performed extensive numerical simulations of the
dynamics for large $N$. A standard
procedure for simulations is to integrate the equation of motion
(\ref{EOM}) for each of the $N$ particles; this involves computing at every integration step a sum
over $N$ terms for each particle, thus requiring a total
computation time scaling with $N$ as $N^2$. In order to perform faster
simulations, we adopted the algorithm discussed in Ref. \cite{Gupta:2012} to compute the sum in
Eq. (\ref{EOM}). This algorithm requires a total computation time that scales with $N$ as
$N \ln N$. The equations of motion were integrated using a
fourth-order Runge-Kutta algorithm with time step equal to $0.01$. The
initial state of the system was chosen to be the uniform state
(\ref{uniform-state}), prepared by having each particle degree of
freedom $\th$ uniformly distributed in $[-\pi,\pi]$, independently of the
others. We report here simulation results for system sizes $2^{14}$ and $2^8$.

In simulations, we monitor the observable
\be
r_m(t)=\fr{1}{N}\Big|\sum_{j=1}^N e^{i(\th_j+2\pi jm/N)}\Big|;
m=0,1,2,\ldots;
\ee
in particular, $r_0(t)$ does not contain any spatial dependence, and, thus, characterizes the
mean-field mode. In the continuum limit, we have
\be
r_m(t)=\Big|\int d\th ds ~e^{i(\th+2\pi ms)}\rho(\th;s,t) \Big|.
\l{rm-defn}
\ee

Starting at time $t=0$ from the uniform state such that
$r_m(0)=0 ~\forall~ m$, and at a temperature $T =0.01$ with
$\alpha=0.25$, we expect from Table \ref{table-Tc} that the Fourier
modes $0, 1, 2, 3$ in particular are 
linearly unstable. Consequently, $r_0(t),r_1(t),r_2(t),r_3(t)$ are expected
to grow in time. Indeed, Fig. \ref{alpha-T-observables}(a) displaying the simulation results
for $r_0(t),r_1(t),r_2(t),r_3(t)$ does show that they all grow in time. However, in the long-time limit, we see that $r_0(t)$
saturates to a value very close to unity, while $r_1(t),r_2(t),r_3(t)$ all decay to
a value very close to zero. By repeating simulations at different
temperatures $T <1/2$ for different values of $\alpha < 1$ (see Fig.
\ref{alpha-T-observables}(b),(c)), we have confirmed that $r_0(t)$ always saturates
in the long-time limit to a non-zero stationary state value
$r_0^{st}=r_0^{st}(T)$ that depends on
the temperature, while $r_m(t)$'s, with $m \ne 0$, decay in the long-time
limit to a stationary state value very close to zero. As shown in
Fig. \ref{scaling-fig}, the
asymptotic value of $r_0(t)$ does not scale with the system size $N$,
while that of $r_m(t)$ for $m \ne 0$ scales with the system size as
$1/\sqrt{N}$, and hence approaches zero in the continuum limit $N \to
\infty$. These observations
suggest that for temperatures $T <1/2$ for all values of $\alpha < 1$, the
stationary state
is characterized by a distribution $\rho_0$ which is non-uniform in
$\th$, but uniform in $s$, and may be obtained from Eq.
(\ref{nonuniform-rho}) as:
 \bea
\rho_0(\th)&\propto&\exp\Big[\fr{\kappa(\a)}{T}\int
d\th'ds'\fr{\cos(\th'-\th)}{(|s'-s|_c)^{\alpha}}\rho_0(\th')\Big]\nonumber
\\
&=&\exp\Big[\fr{1}{T}(m_x \cos \th+m_y\sin \th)\Big],
\l{nonuniform-rho-2}
\eea
where 
\be
(m_x,m_y)\equiv\int_0^{2\pi}d\th (\cos \th,\sin \th) \rho_0(\th).
\l{mxmy}
\ee

\section{Non-uniform stationary state and dynamics of Fourier modes}
\l{nonuniform}
Our observations regarding saturation of $r_0(t)$ and decay of
$r_m(t)$'s, with $m \ne 0$, detailed in the preceding section, raise an
immediate question: Can we show by performing a linear stability
analysis of the state (\ref{nonuniform-rho-2}) at temperatures $T<1/2$
that all the modes of fluctuations decay in the long-time limit to zero,
so that the state is linearly stable for all values of $\alpha < 1$?

Before proceeding with the stability analysis, we note that our system has full rotational invariance in $\th$ at each space point, so
that in the following, we take $m_y=0$ and $ 0 \le m_x \le 1$ without any loss of generality; then, we have
\bea
\rho_0(\th)&=&A\exp\Big[\fr{1}{T}m_x \cos \th\Big],
\l{nonuniform-rho-1}
\eea
where $A$ is the normalization: 
\be
A=\fr{1}{\int_0^{2\pi} d\th e^{\fr{m_x}{T}
\cos \th}}.
\l{normfact}
\ee
Note that the quantity $m_x$ is determined self-consistently by combining
Eq. (\ref{nonuniform-rho-1}) with  Eq. (\ref{mxmy}); one gets
\be
m_x=\fr{I_1(m_x/T)}{I_0(m_x/T)},
\l{mx-eqlbm}
\ee
where $I_n(x)$ is the modified Bessel function of order $n$:
\be
I_n(x) = \fr{1}{2\pi}\int_0^{2\pi} d\th ~e^{x \cos \th}\cos n\th.
\l{besseldef}
\ee
We thus have $A = \left[ 2\pi I_0(m_x/T)\right]^{-1}$.

To study the linear stability of the state (\ref{nonuniform-rho-1}), we
write
\be
\rho(\th;s,t)=\rho_0(\th)+\delta \rho(\th; s,t); ~~~~\delta \rho(\th; s,t) \ll
1,
\l{rho-expansion}
\ee
where we expand $\delta \rho(\th; s,t)$ as
\be
\delta \rho(\th; s,t)=\sum_{m=-\infty}^\infty \widehat{\delta
\rho}_m(\th,t) e^{i2\pi ms}.
\l{rho-expansion-1}
\ee
Using Eq. (\ref{rho-expansion-1}) in Eq. (\ref{FP-equation}), we find
that $\widehat{\delta \rho}_m(\th,t)$ satisfies at its leading order the
equation
\bea
&&\fr{\partial \widehat{\delta\rho}_m}{\partial
t}=m_x\fr{\partial (\sin \th ~\widehat{\delta \rho}_m)}{\partial \th}
+T\fr{\partial^2\widehat{\delta \rho}_m}{\partial
\th^2}\nonumber \\
&&-\lambda_m(\a)\fr{\partial }{\partial \th}\Big(\Big[\int
d\th'\sin(\th'-\th)\widehat{\delta
\rho}_m(\th',t)\Big]
\rho_0\Big), 
\l{FP-equation-linear-2}
\eea
where we have defined
\be
\lambda_m(\a)\equiv\kappa(\a)\Lambda_m(\a).
\ee

The eigenfrequencies associated with the linear equation
(\ref{FP-equation-linear-2}) can be studied by looking for solutions of the form
\be
\widehat{\delta \rho}_m(\th,t)=\widetilde{\delta
\rho}_m(\th,\omega)e^{i\omega t}.
\l{longtime}
\ee
Before studying the spectrum of the eigenfrequency $\omega$, it is instructive
to analyze the explicit solution $\widetilde{\delta \rho}_m(\th,0)$ corresponding to the neutral mode $\omega = 0$. We find from Eq.
(\ref{FP-equation-linear-2}) that $\widetilde{\delta \rho}_m(\th,0)$
satisfies
\bea
&&m_x \sin \th ~\widetilde{\delta \rho}_m-\lambda_m(\a)\Big(\widetilde{m}_y \cos \th- \widetilde{m}_x \sin
\th\Big)
\rho_0\nonumber \\
&&+T\fr{\partial\widetilde{\delta \rho}_m}{\partial
\th}=C, 
\l{FP-equation-linear-5}
\eea
where $C$ is a constant independent of $\th$, and
\bea
(\widetilde{m}_x,\widetilde{m}_y)=\int d\th (\cos \th,\sin \th)
\widetilde{\delta \rho}_m(\th,0).
\l{widetildemxmy}
\eea

Equation (\ref{FP-equation-linear-5}) has the solution
\bea
\widetilde{\delta \rho}_m(\th,0)&=&\widetilde{\delta
\rho}_m(0,0)e^{-\fr{m_x}{T}(1-\cos \th)} \nonumber \\
&&+\fr{C}{T}e^{\fr{m_x}{T}\cos \th} \int_0^\th d\th' e^{-\fr{m_x}{T}\cos
\th'}\nonumber \\
&&+\fr{A\lambda_m(\alpha)\widetilde{m}_y}{T} e^{\fr{m_x}{T}\cos \th}
\sin \th \nonumber \\
&&-\fr{A\lambda_m(\alpha)\widetilde{m}_x}{T} e^{\fr{m_x}{T}\cos \th}
(1-\cos \th). 
\l{deltarho} 
\eea
The periodicity condition $\widetilde{\delta \rho}_m(\th+2\pi,0)=\widetilde{\delta
\rho}_m(\th,0)$ implies that $C=0$. Normalization of $\rho(\th;s,t)$ for all times implies, using Eq.
(\ref{rho-expansion}) and the fact that $\rho_0(\th)$ is normalized, that
$\int_0^{2\pi} d\th ~\widetilde{\delta \rho}_m(\th,0)=0 ~ \forall ~m$.
Then, from Eq. (\ref{deltarho}) with $C=0$, we get
\bea
&&\widetilde{\delta
\rho}_m(0,0)\int_0^{2\pi} d\th e^{-\fr{m_x}{T}(1-\cos \th)}\nonumber \\
&&=\fr{A\lambda_m(\alpha)\widetilde{m}_x}{T} \int_0^{2\pi} d\th e^{\fr{m_x}{T}\cos \th}
(1-\cos \th).  
\l{norm}
\eea
Using the above equation and Eq. (\ref{mx-eqlbm}) in Eq.
(\ref{deltarho}) with $C=0$, we arrive at the following
expression for
$\widetilde{\delta \rho}_m(\th,0)$:
\bea
&&\widetilde{\delta \rho}_m(\th,0) \nonumber \\
&&=\fr{A\lambda_m(\alpha)}{T} e^{\fr{m_x}{T}\cos \th} \left[\widetilde{m}_x
\left( \cos \th - m_x \right) + \widetilde{m}_y \sin \th \right].
\l{deltarho_bis} 
\eea

\begin{figure}[h!]
\includegraphics[width=75mm]{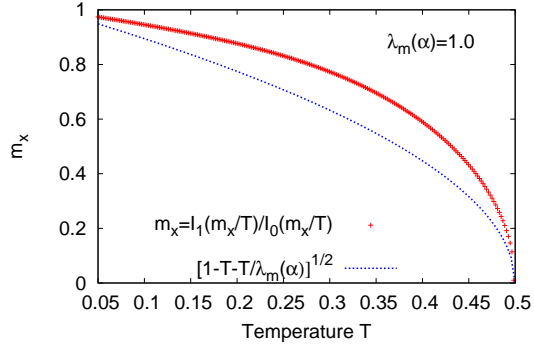}
\caption{(Color online) Plot showing as a function of $T$ the values of
$m_x$ that satisfy Eq. (\ref{mx-eqlbm}), and those that satisfy
(\ref{secondcond_bis}) at $\lambda_m(\alpha)=1$. The fact that the
curves do not intersect at any $T$ in the range $0<T<1/2$ shows that in
this range, the value of $m_x$
that solves Eq. (\ref{mx-eqlbm}) at a given $T$ does not satisfy Eq.
(\ref{secondcond_bis}).}
\label{consistency-plot}
\end{figure}

The quantities $\widetilde{m}_{x,y}$ in Eq. (\ref{deltarho_bis}) are
determined self-consistently. They cannot both be equal to zero, as would have happened if $\widetilde{\delta \rho}_m(\th,0)$ were a
constant, but this is not possible as the integral of $\widetilde{\delta
\rho}_m(\th,0)$ from $0$ to $2\pi$ is required to vanish. 
Using Eqs. (\ref{widetildemxmy}), (\ref{deltarho_bis}), (\ref{normfact}) and (\ref{mx-eqlbm}), we get
the following two equations:
\bea
&&\widetilde{m}_y = \widetilde{m}_y \fr{\lambda_m(\alpha)}{m_x}\fr{I_1(m_x/T)}
{I_0(m_x/T)}, \l{firstcond}\\
&&\widetilde{m}_x = \widetilde{m}_x \fr{\lambda_m(\alpha)}{T} \left(
1 - T - m_x^2 \right) \l{secondcond},
\eea
where $0<T<1/2$, and $m_x$ satisfies Eq. (\ref{mx-eqlbm}). Equation (\ref{firstcond})
is obviously satisfied for $\widetilde{m}_y = 0$. If $\widetilde{m}_y \ne 0$,
then, using Eq. (\ref{mx-eqlbm}), we see that Eq. (\ref{firstcond}) is an
identity for $m=0$, while it has no solution for $m \ne 0$, which follows from
the property that $\lambda_m(\alpha) < 1$ for $m \ne 0$. 
Let us now analyze Eq. (\ref{secondcond}). It is satisfied for
$\widetilde{m}_x = 0$. If $\widetilde{m}_x \ne 0$, it gives
\be
m_x = \sqrt{1 - T - T/\lambda_m(\alpha)},
\l{secondcond_bis}
\ee
giving a relation between $m_x$ and $T$, which has to be satisfied together with Eq.
(\ref{mx-eqlbm}). To check whether the two equations are consistent,
note that $\lambda_m(\alpha)=\lambda_{-m}(\alpha)$ and that
$\lambda_m(\alpha) \ge 0$ is a monotonically
decreasing function of $|m|$, with $\lambda_0(\alpha)=1$ and $\lim_{m \to
\pm \infty}\lambda_m(\alpha)=0$. Hence, to check the consistency, we have to consider in  Eq. (\ref{secondcond_bis}) 
any value for $\lambda_m(\alpha)$ between $0$ and $1$.
Actually, when $\lambda_m(\alpha) < 1$, Eq. (\ref{secondcond_bis}) gives
a real value for $m_x$ only for $T$ in the range
$0<T<\fr{\lambda_m(\alpha)}{1+\lambda_m(\alpha)}$, which is smaller than
the range $0<T<1/2$.
In Fig. \ref{consistency-plot}, we plot $m_x(T)$ as given by Eq.
(\ref{secondcond_bis}) for $\lambda_m(\alpha)=1$, and as given by
solving the implicit equation (\ref{mx-eqlbm}). We see that the two curves
do not intersect. This also implies that there cannot be any intersection
for $\lambda_m(\alpha) <1$, since the right hand side of Eq.
(\ref{secondcond_bis}) decreases on decreasing $\lambda_m(\alpha)$.
Thus, we conclude that Eq.
(\ref{secondcond_bis}) for $0 < T < 1/2$ is not solved by $m_x$
that satisfies Eq. (\ref{mx-eqlbm}). In summary, the two equations, 
(\ref{firstcond}) and (\ref{secondcond}), are not satisfied by $m_x$
given by Eq. (\ref{mx-eqlbm}), when $m\ne0$. When $m=0$, there is only one
solution, namely, $\widetilde{m}_x = 0$, and $\widetilde{m}_y$ any
non-zero value. For this solution, on using Eqs.
(\ref{nonuniform-rho-1}) and (\ref{deltarho_bis}), we have
\be
\widetilde{\delta \rho}_0(\th,0)= \fr{1}{T} \rho_0 (\th) \widetilde{m}_y
\sin \th,
\l{deltarho_ter} 
\ee
which is just a perturbation of $\rho_0(\th)$ that corresponds to a
rotation in $\th$ space of all the rotators of the system. It is therefore natural that it corresponds
to a neutral mode.
\begin{figure}[here!]
\centering
\includegraphics[width=75mm]{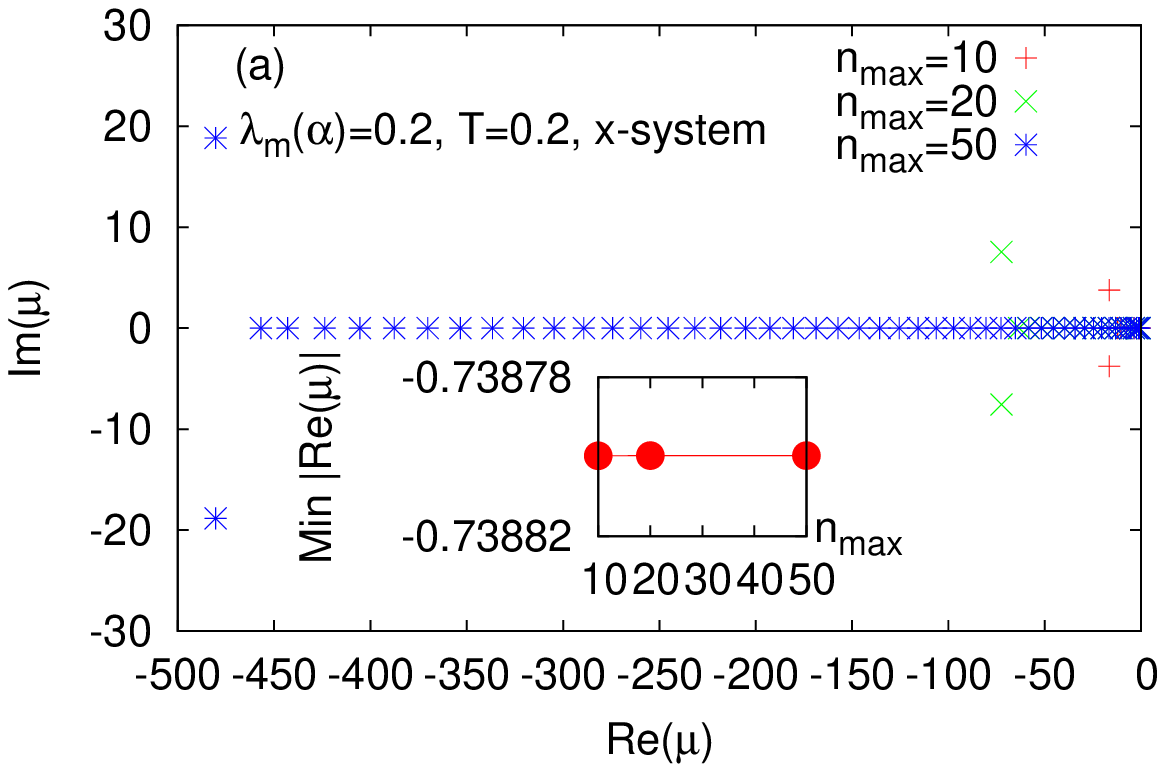} \\
\includegraphics[width=75mm]{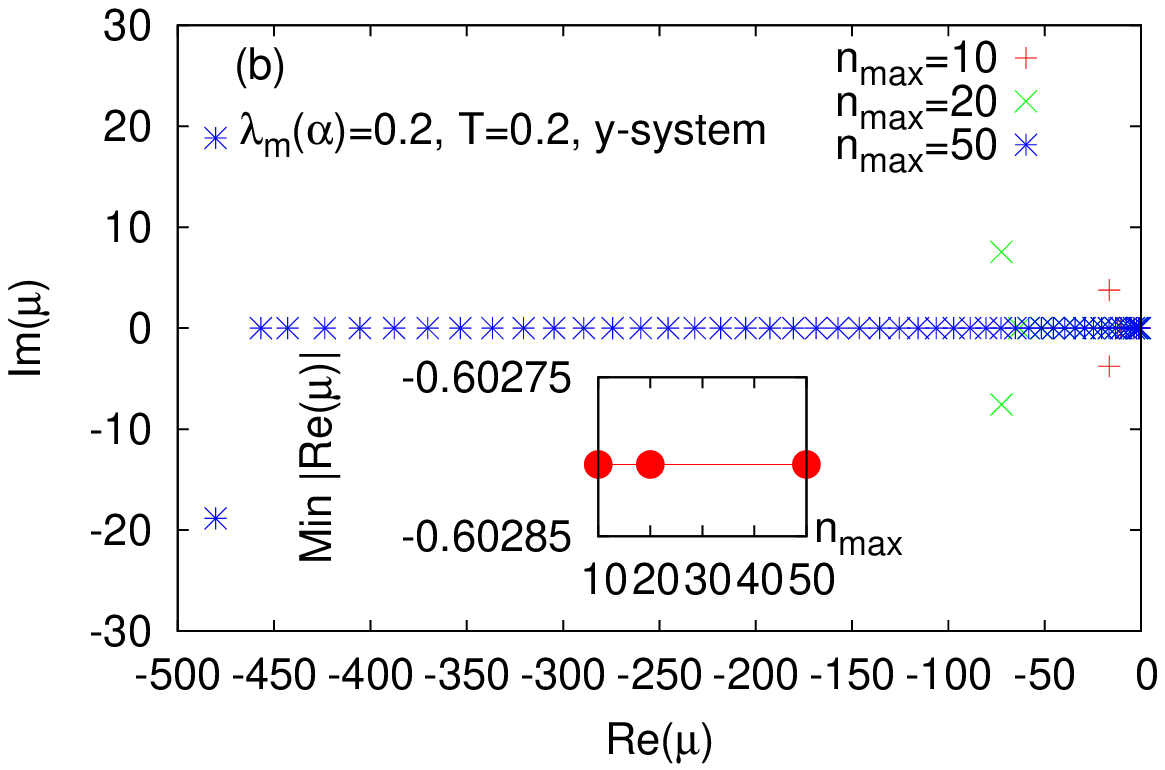}\\
\includegraphics[width=75mm]{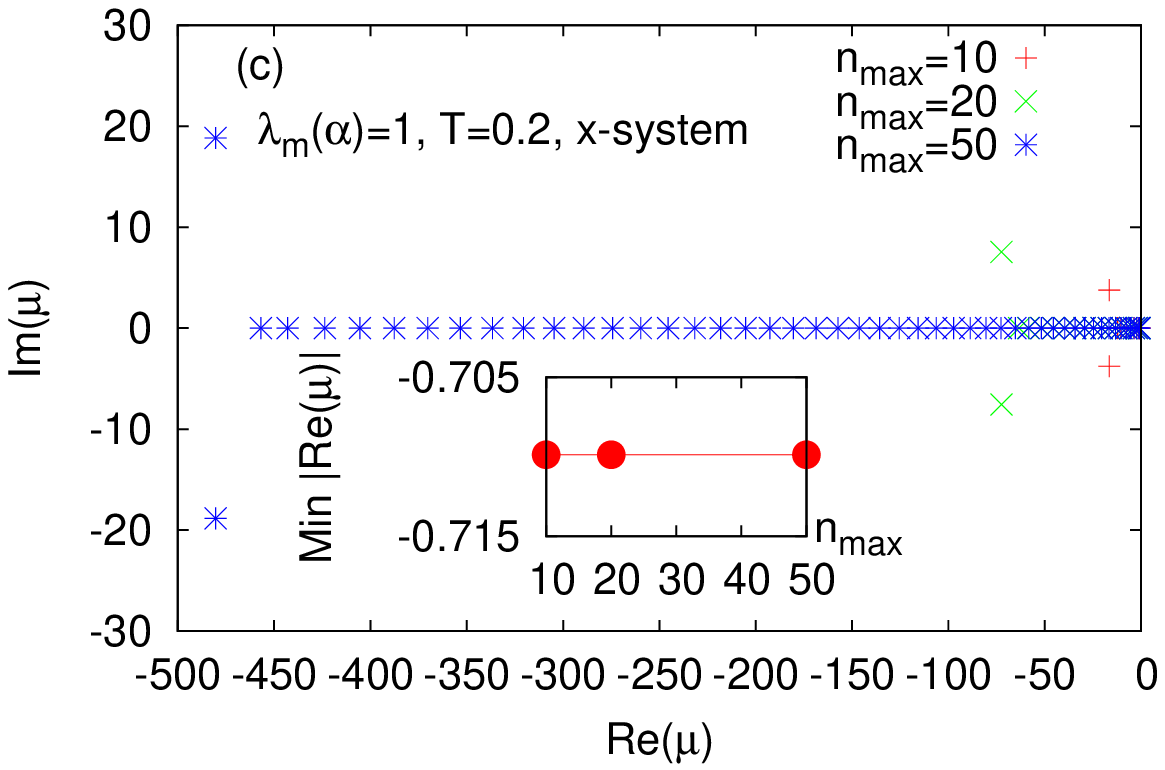}\\
\includegraphics[width=75mm]{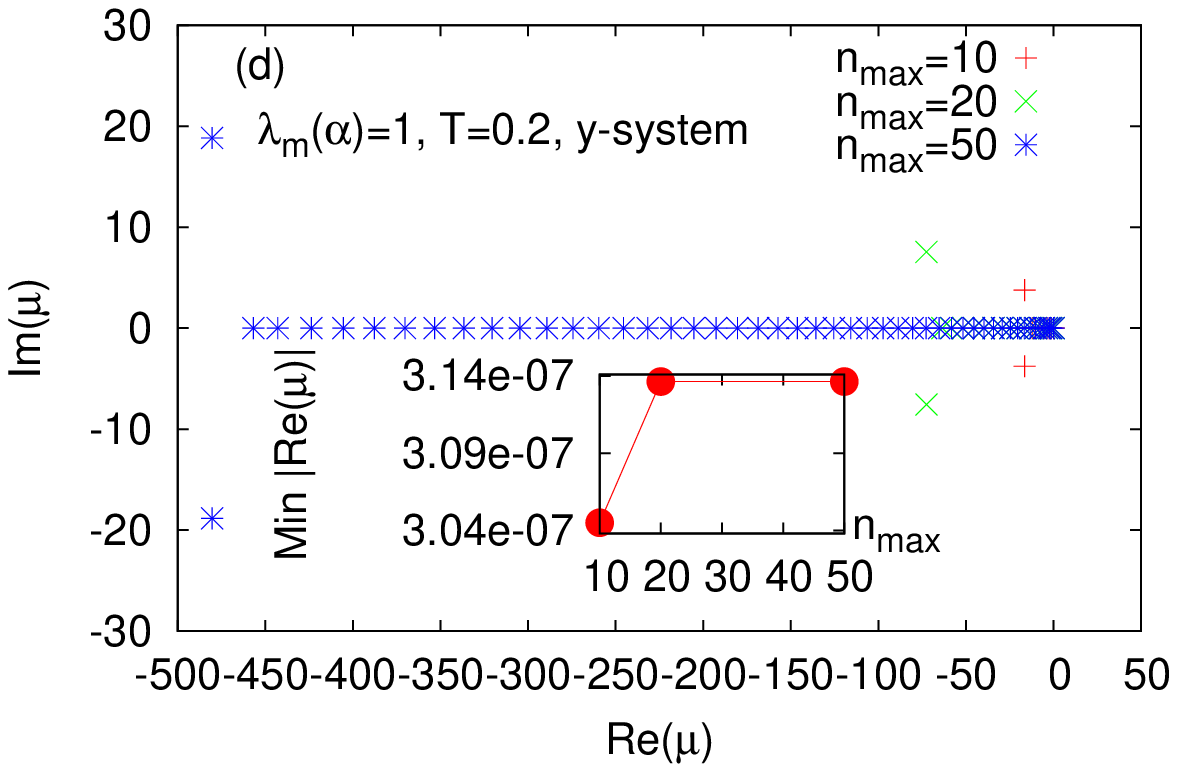}
\caption{(Color online) Real and imaginary parts of the eigenvalues $\mu$
of the $x$-system, Eq. (\ref{xequation}), and the $y$-system, Eq.
(\ref{yequation}), for
a $m \ne 0$ mode such that $\lambda_m(\alpha)<1$ and for the $m=0$ mode such
that $\lambda_m(\alpha)=1$. The temperature is $T=0.2$. The parameter
$n_{max}$ denotes the order of truncation of the eigenvalue equations.}
\label{eigenvalues-0.2}
\end{figure}

On the basis of the above discussion, we conclude that there is no temperature $T<1/2$
such that the $m$th mode of fluctuations, for any $m$, has a vanishing
eigenfrequency, excepting for the trivial one for $m=0$ that
corresponds to a state
obtained by a rotation in $\th$ space of all the rotators of the system.
Thus, the $m$th mode either grows or
decays in time at all temperatures $T <1/2$, corresponding to
eigenfrequencies $\omega$ which have respectively a negative or a
positive imaginary part. 

In order to examine the entire spectrum of the eigenfrequency $\omega$, we
use Eqs. (\ref{FP-equation-linear-2}) and (\ref{longtime}),
and the substitution $\mu \equiv i\omega$ to get the following equation
for $\widetilde{\delta \rho}_m(\th,\omega)$.
\bea
&&\mu \widetilde{\delta\rho}_m=m_x\fr{\partial (\sin \th ~\widetilde{\delta \rho}_m)}
{\partial \th}+T\fr{\partial^2\widetilde{\delta \rho}_m}{\partial
\th^2}
\nonumber \\
&&-\lambda_m(\a)\fr{\partial }{\partial \th}\Big(\Big[\int
d\th'\sin(\th'-\th)\widetilde{\delta
\rho}_m(\th',\omega)\Big]
\rho_0\Big). 
\l{FP-equation-linear-3_bis}
\eea
Let us define
\bea
&&(\widetilde{m}^{(n)}_x,\widetilde{m}^{(n)}_y)=\int d\th (\cos n\th,\sin
n\th)\widetilde{\delta \rho}_m(\th,\omega), \nonumber \\
\l{widetildemxmyn} \\
&&m^{(n)}_x=\int d\th \cos n\th ~\rho_0(\th)=\fr{I_n(m_x/T)}{I_0(m_x/T)}, \nonumber
\eea
where we note that $m^{(1)}_x=m_x$. On multiplying Eq. (\ref{FP-equation-linear-3_bis}) in turn by $\cos n\th$ and
$\sin n\th$, integrating over $\th$ from $0$ to $2\pi$, noting that
$\rho_0$ is even in $\th$, that $\widetilde{\delta \rho}_m(2\pi,\omega)=\widetilde{\delta
\rho}_m(0,\omega)$, and that $\rho_0(2\pi)=\rho_0(0)$, we arrive at the following system
of equations for $n \ge 1$:
\bea
\mu \widetilde{m}^{(n)}_x &=& \frac{1}{2}nm_x \Big(\widetilde{m}^{(n-1)}_x - \widetilde{m}^{(n+1)}_x
\Big) - T n^2 \widetilde{m}^{(n)}_x \nonumber \\
&&+ \frac{1}{2}\lambda_m(\alpha)n \widetilde{m}^{(1)}_x \Big(m^{(n-1)}_x
- m^{(n+1)}_x \Big), \l{xequation} \\
\mu \widetilde{m}^{(n)}_y &=& \frac{1}{2}nm_x \Big(\widetilde{m}^{(n-1)}_y - \widetilde{m}^{(n+1)}_y
\Big) - T n^2 \widetilde{m}^{(n)}_y \nonumber \\
&&+ \frac{1}{2}\lambda_m(\alpha)n \widetilde{m}^{(1)}_y \Big(
m^{(n-1)}_x + m^{(n+1)}_x \Big), \l{yequation}
\eea
where
$\widetilde{m}^{(0)}_x=\widetilde{m}^{(0)}_y=0$, and $m^{(0)}_x=1$.
The operators on the right-hand sides of the above equations being
non-hermitian have in general both real and complex eigenvalues.
\begin{figure}[here!]
\includegraphics[width=75mm]{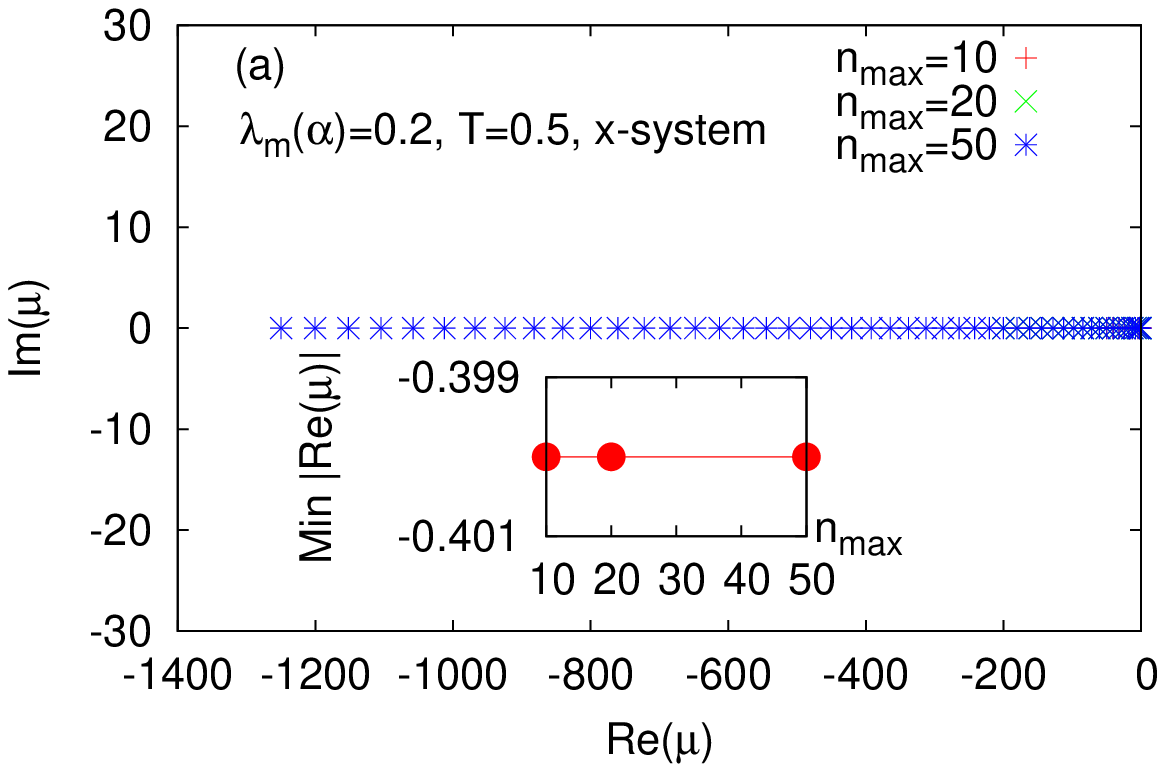}\\
\includegraphics[width=75mm]{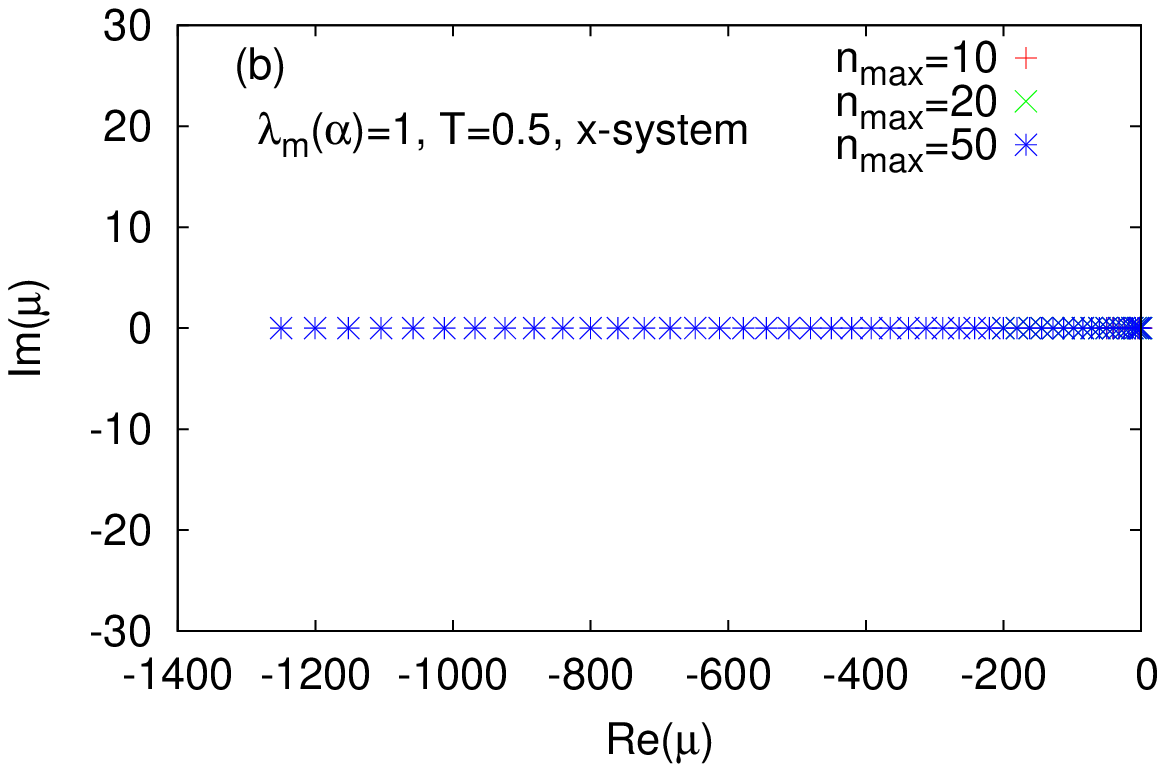}
\caption{(Color online) Real and imaginary parts of the exact eigenvalues $\mu$
of the $x$-system, Eqs. (\ref{xequation-1}) and (\ref{xequation-2}), for
a $m\ne0$ mode such that $\lambda_m(\alpha)<1$ and for the $m = 0$ mode such
that $\lambda_m(\alpha)=1$. The temperature is $T=0.5$. The parameter
$n_{max}$ denotes the order of truncation of the eigenvalue equations.}
\label{eigenvalues-0.5}
\end{figure}

When $T < 1/2$ and $m^{(n)}_x \ne 0$ for all $n \in \mathcal{N}$, then,
combined with Eq. (\ref{mx-eqlbm}), the eigenvalues of the system of
equations, (\ref{xequation}) and (\ref{yequation}), can be obtained numerically after truncating the equations
to a finite value of $n$, say, $n_{max}$. As argued before on the basis of properties of
$\lambda_m(\alpha)$, in order to probe the behavior of a $m \ne 0$
mode of fluctuations, it suffices to consider any real positive value
$<1$ for $\lambda_m(\alpha)$, while for the $m=0$ mode, we have
$\lambda_m(\alpha)=1$.  The results for the eigenvalue spectrum of Eqs.
(\ref{xequation}) and (\ref{yequation}) are shown in Fig.
\ref{eigenvalues-0.2} for the temperature
$T=0.2$. 

When $T \ge1/2$ and $m^{(n)}_x=0$ for all $n \in \mathcal{N}$, the eigenvalue equations,
(\ref{xequation}) and (\ref{yequation}), simplify for $n \ge 2$ to  
\bea
\mu \widetilde{m}^{(n)}_x &=& - T n^2 \widetilde{m}^{(n)}_x, \l{xequation-1} \\
\mu \widetilde{m}^{(n)}_y &=& - T n^2 \widetilde{m}^{(n)}_y, \l{yequation-1}
\eea
while for $n=1$, the equations are 
\bea
\mu \widetilde{m}^{(1)}_x &=& - T \widetilde{m}^{(1)}_x + \frac{1}{2}\lambda_m(\alpha) \widetilde{m}^{(1)}_x, \l{xequation-2} \\
\mu \widetilde{m}^{(1)}_y &=& - T \widetilde{m}^{(1)}_y + \frac{1}{2}\lambda_m(\alpha)\widetilde{m}^{(1)}_y. \l{yequation-2}
\eea
The system of equations, (\ref{xequation-1})-(\ref{yequation-2}) is a set of
independent equations, and may be
solved quite easily to get the exact eigenvalues $\mu$; from the
equations, it is clear that these
eigenvalues are the same for the $x$ and the $y$-system. Of course, the number of eigenvalues depends on the value $n_{max}$ of $n$ at
which the equations are truncated. In particular,  for
$\lambda_m(\alpha)=1$ and $T=1/2$, one finds the eigenvalue $\mu=0$.
The results for the eigenvalue spectrum of Eqs.
(\ref{xequation-1})-(\ref{yequation-2}) are shown in Fig. \ref{eigenvalues-0.5} for the temperature $T=1/2$.

Let us first discuss the results displayed in Fig.
\ref{eigenvalues-0.2}. From the $m \ne 0$ mode results, displayed in
panels (a) and (b), we see that the
eigenvalues of the $x$- and the $y$-system are both real and complex, but
the important thing to note is that the real eigenvalues are all
negative while the complex ones have strictly negative real parts. In
order to demonstrate the former fact, a zoom into the region near the zero
of the $Re(\mu)$ axis, shown in the insets, illustrates that the
eigenvalue closest to 0 of both the $x$- and the $y$-system has
a negative real part, so that the remaining eigenvalues having larger real
parts in magnitude are thus all negative. The insets also show that the
eigenvalue with the smallest negative real part converges in magnitude on increasing the
truncation order $n_{max}$. For higher $n_{max}$, only the number of
eigenvalues increases; in particular, only the number of real
eigenvalues with larger magnitude increases, while there is still only one complex eigenvalue
which has the largest negative real part. Computing the eigenvalues for
other $\lambda_m$'s and temperatures $T <1/2$, we find that the number
of complex eigenvalues is always small (with the number increasing for
smaller $\lambda_m$'s and $T$'s), and these eigenvalues always have large negative
real parts. Since the $m$th mode of fluctuations has
the time dependence $\sim e^{\mu t}$ (recall Eq. (\ref{longtime}) and the
definition of $\mu=i\omega$), and since all values for $\mu$ have
strictly negative real parts, it follows that at temperatures $T <1/2$, the $m$th mode
of fluctuations, with $m \ne 0$, decays in the long-time limit to zero. 
Figure \ref{eigenvalues-0.2}, panel (c) shows that the behavior for
the $m=0$ mode is very similar to that observed in panels (a) and (b)
for $m \ne 0$, discussed above. Panel (d) also has the same general
behavior, excepting that now the eigenvalue closest to the origin of the
$Re(\mu)$ axis, shown in the inset as a function of $n_{max}$, is very
slightly ($\sim 10^{-7}$) positive. Actually, this eigenvalue should be
exactly zero, being the one corresponding to rotational invariance that
we discussed earlier (see the discussion following Eq.
(\ref{secondcond_bis})). In fact, using Eq. (\ref{deltarho_ter}) in Eq.
(\ref{widetildemxmyn}) to compute $\widetilde{m}^{(n)}_y$, and using it
in Eq. (\ref{yequation}), it may be checked that its right hand side is zero for any $n$, that is, the particular
solution (\ref{deltarho_ter}) gives $\mu = 0$. The slight positive value is just a numerical
artifact.

We now discuss the results displayed in Fig.
\ref{eigenvalues-0.5}. We show here only the eigenvalues of the $x$-system, since, as noted earlier, the $y$-system has exactly the same eigenvalue
spectrum. From the $m \ne 0$ mode results, displayed in
panel (a), we see that the
eigenvalues are only real and negative,
with no complex eigenvalue. The eigenvalue closest to the origin of the
$Re(\mu)$ axis, shown in the inset as a function of $n_{max}$, dominates the
behavior of the $m$th mode of fluctuations in time, thereby making it
decay to zero in the long-time limit. From the $m = 0$ mode results, displayed in
panel (b), we see that the eigenvalues are real and negative. As already mentioned above, the
eigenvalue closest to the origin is precisely zero, which makes the zero mode of fluctuations to neither grow nor decay at the
temperature $T=1/2$. 
\begin{figure}[here!]
\centering
\includegraphics[width=75mm]{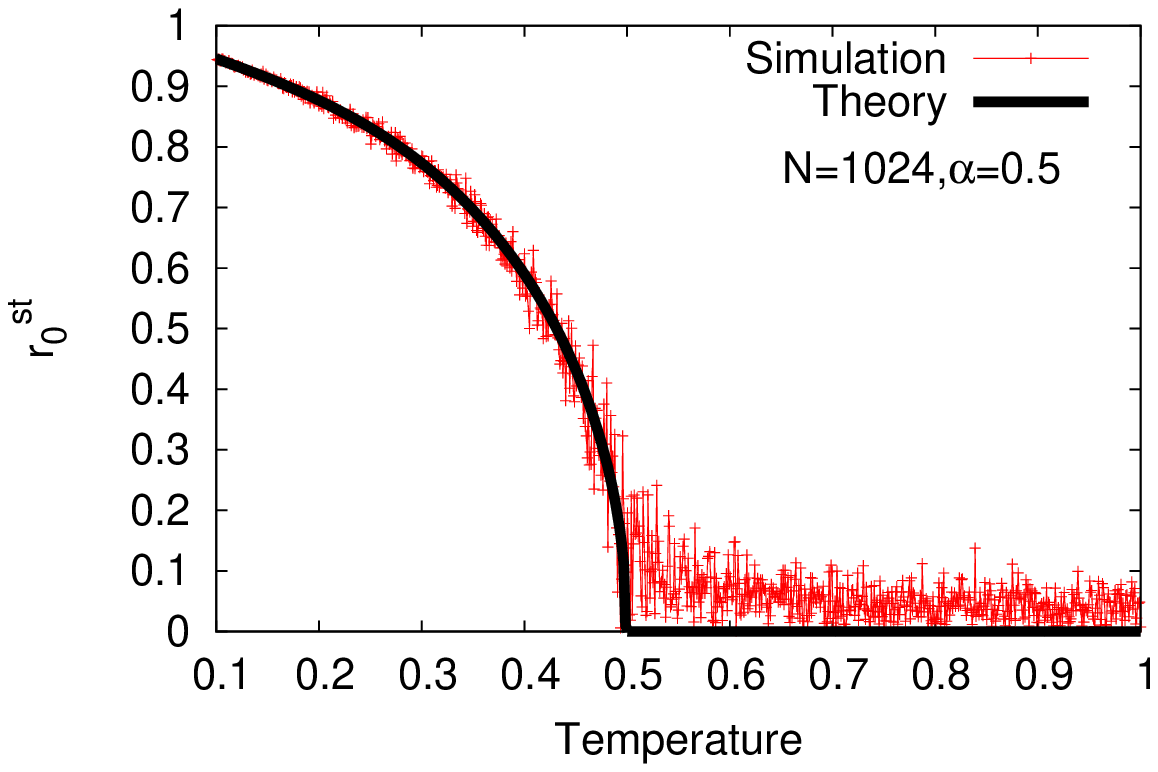}
\caption{Plot showing $r_0^{st}$ as a function of temperature. The data are obtained
by starting from the uniform state (\ref{uniform-state}) at a high
temperature $T=1$, and then changing adiabatically the temperature as a
function of time $t$ as $T(t)=1-Rt$, where the rate $R$ is taken to be
such that $R\tau_{st} \ll 1$, with $\tau_{st}$ being the time to reach
stationary
state at a fixed temperature. The condition $R\tau_{st} \ll 1$ ensures that between two successive
values of $T$, the system gets enough time to attain stationarity. For a system of
size $N$, taking $\tau_{st} = N^2dt$, where $dt$ is the numerical
integration timestep, we take $R=0.1/N^2dt$.}
\l{eqlbm-r0}
\end{figure}

Based on results discussed in this section and in section
\ref{uniform}, we conclude that the linearly stable stationary state of the dynamics
(\ref{EOM}) in the $N \to \infty$ limit is a state uniform in space but
non-uniform in $\th$ when the temperature is less that $1/2$. 
On tuning
the temperature to a value larger than $1/2$, the stationary state becomes
uniform in $\th$. Thus, our linear stability analysis in the limit $N
\to \infty$ allows us to predict that the system undergoes a transition at the temperature
$T=T_{c,0}=1/2$. Indeed, simulation results shown in Fig.
\ref{eqlbm-r0} for $r_0^{st}$ suggests a continuous phase transition as a function of temperature, where we also show the theoretical curve for $r_0^{st}$
obtained by combining Eq. (\ref{rm-defn}) with Eq. (\ref{nonuniform-rho-1}) to give for $T
<1/2$ the result
\be
r_0^{st}=\fr{I_1(m_x/T)}{I_0(m_x/T)}=m_x;
\ee
for $T>1/2$, on using Eq. (\ref{uniform-state}), one has
\be
r_0^{st}=0.
\ee
Exactly at $T=1/2$, we have $r_0^{st}=0$.
\section{Conclusions and perspectives}
In this paper, we considered a paradigmatic long-range interacting model of
particles residing on the sites of a one-dimensional periodic lattice.
Each particle has an internal degree of freedom $\th$ which is coupled to
those of other particles with an attractive $XY$-like interaction $\sim -\cos(\th_i-\th_j)$ between the $i$th and $j$th particles, with
the coupling strength decaying with the interparticle separation $r$ as
$1/r^\alpha$; ~$0 < \alpha < 1$. We considered the overdamped dynamics within a canonical
ensemble of this so-called $\alpha$-HMF model. We studied the model
numerically, and also analytically in the continuum limit through the Fokker-Planck equation for the evolution of the local
density of particles. 

The Fokker-Planck equation allows a stationary state which is
uniform in both $\th$ and space $s$ of the lattice. A linear stability
analysis of such a state shows that it is stable above the temperature
$T=1/2$. Below this temperature, when it is unstable, numerics as well as the linear stability
analysis show that different spatial Fourier modes of fluctuations (the number of which depends on the
temperature) grow exponentially in time with
different rates. However, our numerical simulations also show that all 
the non-zero Fourier modes decay at long times to zero. By contrast, the
zero mode (the ``mean-field" mode) grows and
reaches a non-zero value, corresponding to a non-uniform stationary
state, i.e. a state which is non-uniform in $\th$, but uniform in $s$.

On the basis of our investigations, it appears that both for temperatures below and above $T=1/2$, the state which is uniform in space acts as a global
attractor of all possible stationary states of the $\alpha$-HMF model. The
state is the same as the equilibrium state of the mean-field version of
the model, the HMF model. For temperatures below $1/2$, the late-time
damping of the unstable non-zero modes after their initial growth
is an interesting and unexpected phenomenon. For such
temperatures, by developing a linear stability analysis
around the non-uniform state, we showed that
all fluctuations around that state decay in time, thereby stabilizing
the state; describing analytically the complete evolution of the
unstable non-zero modes, from
their initial growth in time to their late-time decay, thereby leading
to the convergence to the non-uniform state remains an interesting challenging problem. 

The model we studied may be related to the Kuramoto model in the field of non-linear
dynamical systems \cite{Kuramoto:1984,Strogatz:2000,Acebron:2005}. Interpreting each
of the $N$ particles of our system to be a phase-only
oscillator with phase $\th_i$,
the equation of motion (\ref{EOM}) may be looked upon as the one
governing the time evolution of the phase of the
$i$th oscillator, in the presence of thermal noise. On considering
$\alpha=0$ in Eq. (\ref{EOM}), the model in the absence of thermal
noise is a particular limit of the Kuramoto model in which each
oscillator has the same intrinsic frequency of
oscillation. In the presence of thermal
noise, Eq. (\ref{EOM}) with $\alpha=0$ is the generalization of the
Kuramoto model considered in Ref. \cite{Sakaguchi:1988}. On considering
$\alpha \ne 0$, and attributing to each
oscillator an intrinsic frequency sampled from a distribution, the model in the absence of
thermal noise has been considered in
\cite{Rogers:1996,Chowdhury:2010,Uchida:2011,Gupta:2012}. In particular, a
mean-field dominance similar to the one observed in this work has been
seen in numerics in Ref. \cite{Gupta:2012}. It remains an open problem
to relate this dominance in these two scenarios.  

\section{Acknowledgments}
S. G. and S. R. acknowledge the contract ANR-10-CEXC-010-01 for support.
The motivation for this work arose from discussions in two
workshops held at the Korea Institute for Advanced Study (KIAS), Seoul, and at the
Centre Blaise Pascal, ENS-Lyon in July and August, 2012, respectively.
S. G. and S. R. thank KIAS for hospitality during their mutual visit in
July, 2012. A. C. acknowledges ENS-Lyon for hospitality during some
stages of the work presented in this paper. We acknowledge useful discussions with David Mukamel and Hyunggyu Park.

\end{document}